\documentclass[prd,aps,twocolumn,a4paper,floatfix,nofootinbib]{revtex4-1}

\usepackage[utf8]{inputenc}
\usepackage{graphicx,psfrag}
\usepackage{mathrsfs}
\usepackage{amsmath,amsfonts,amssymb}
\usepackage{multirow}
\usepackage{comment}
\usepackage{xcolor}
\usepackage{enumerate}
\usepackage{float}

\usepackage{subfig}
\usepackage{tabularx}

\usepackage[colorlinks=true]{hyperref}
\hypersetup{
  linkcolor=blue,
  citecolor=blue,
  urlcolor=blue
}
\usepackage{bbm}
\usepackage{tensor} %https://www.ctan.org/pkg/tensor
\usepackage{acronym}
\usepackage{xspace}
\usepackage{booktabs,multirow,cellspace}

\usepackage{color}

\definecolor{cyan}{rgb}{0,0.9,0.9}
\definecolor{orange}{rgb}{0.9,0.5,0}
\definecolor{purple}{rgb}{0.8,0.4,0.8}
\definecolor{gray}{rgb}{0.8242,0.8242,0.8242}
\definecolor{pink}{rgb}{1.0, 0.0, 0.5}

\newacro{BBH}{binary-black-hole}
\newacro{BH}{black hole}
\newacro{BNS}{binary neutron star}
\newacro{EOB}{effective-one-body}
\newacro{EOS}{equation-of-state}
\newacro{GW}{gravitational-wave}
\newacro{NR}{numerical-relativity}
\newacro{NS}{neutron star}
\newacro{PN}{post-Newtonian}
\newacro{PSD}{power spectral density}
\newacroplural{PSD}[PSDs]{power spectral densities}
\newacro{SNR}{signal-to-noise ratio}
\newacro{ODE}{ordinary-differential-equation}
\newacro{ZDHP}{zero-detuned high-power configuration}

\def\NRTidal{\texttt{NRTidal}\xspace}

\def\NRtidal2{\texttt{NRTidalv2}\xspace}
\def\SEOBT{\texttt{SEOBNRv4T}\xspace}

\def\SEOBNRROMNRtidal{\texttt{SEOBNR\_ROM\_NRTidalv2}\xspace}
\def\IMRPhenomDNRtidal{\texttt{IMRPhenomD\_NRTidalv1}\xspace}

\def\IMRPhenomP1NRtidal{\texttt{IMRPhenomPv2\_NRTidalv1}\xspace}

\def\IMRPhenomPNRtidal2{\texttt{IMRPhenomPv2\_NRTidalv2}\xspace}
\def\TEOBResumS{\texttt{TEOBResumS}\xspace}
\def\IMRPhenomD{\texttt{IMRPhenomD}\xspace}

\def\IMRPhenomPv2{\texttt{IMRPhenomPv2}\xspace}

\def\Slyuuh{SLy$_{0.57 \uparrow \uparrow}$}
\def\Slyuu{SLy$_{0.37 \uparrow \uparrow}$}
\def\Slyud{SLy$_{0.16 \uparrow \downarrow}$}
\def\Slydd{SLy$_{0.28 \downarrow \downarrow}$}

\definecolor{mygreen}{rgb}{0.1,0.8,0.1}

\usepackage[normalem]{ulem}

\begin{document}

\title{High-accuracy simulations of highly spinning binary neutron star systems}

\author{Reetika \surname{Dudi$^1$}, Tim \surname{Dietrich$^{1,2}$}, Alireza \surname{Rashti$^3$},  Bernd \surname{Br\"ugmann$^4$}, Jan \surname{Steinhoff}$^1$, Wolfgang \surname{Tichy$^3$}}

\affiliation{$^1$Max Planck Institute for Gravitational Physics (Albert Einstein Institute), Am Muhlenberg 1, Potsdam, Germany\\
$^2$ Institut f\"{u}r Physik und Astronomie, Universit\"{a}t Potsdam, D-14476, Potsdam, Germany\\
$^3$ Department of Physics, Florida Atlantic University, Boca Raton, FL 33431 USA\\
$^4$ Theoretical Physics Institute, University of Jena, 07743 Jena, Germany}

\date{\today}

\begin{abstract} 
With an increasing number of expected gravitational-wave detections of binary neutron star mergers, it is essential that gravitational-wave models employed for the analysis of observational data are able to describe generic compact binary systems. This includes systems in which the individual neutron stars are millisecond pulsars for which spin effects become essential. 
In this work, we perform numerical-relativity simulations of binary neutron stars with aligned and anti-aligned spins within a range of dimensionless spins of $\chi \sim [-0.28,0.58]$. 
The simulations are performed with multiple resolutions, show a clear convergence order and, consequently, can be used to test existing waveform approximants. 
We find that for very high spins gravitational-wave models that have been employed for the interpretation of GW170817 and GW190425 are not capable of describing our numerical-relativity dataset. 
We verify through a full parameter estimation study in which clear biases in the estimate of the tidal deformability and effective spin are present. 
We hope that in preparation of the next gravitational-wave observing run of the Advanced LIGO and Advanced Virgo detectors our new set of numerical-relativity data can be used to support future developments of new gravitational-wave models.
\end{abstract}

\maketitle

%%%%%%%%%%%%%%%%%%%%%%%%%%%%%%%%%%%%%%%%%%%%%%%%
\section{Introduction}
\label{Section:Introduction}
%%%%%%%%%%%%%%%%%%%%%%%%%%%%%%%%%%%%%%%%%%%%%%%%

The first detection of gravitational waves (GWs) and 
electromagnetic (EM) signals originating from the 
same astrophysical source, the binary neutron star (BNS) merger
GW170817, has been a scientific breakthrough 
which inaugurated a new era in multi-messenger
astronomy~\cite{TheLIGOScientific:2017qsa,Monitor:2017mdv}.
Followed by this first direct detection of GWs emitted from a BNS 
system, the Advanced LIGO~\cite{TheLIGOScientific:2014jea} and 
Advanced Virgo detectors~\cite{TheVirgo:2014hva} 
observed a second BNS event in April 2019, GW190425~\cite{Abbott:2020uma}. 
In contrast to GW170817, the total mass of GW190425 was larger 
than the mass of BNS systems known to exist in our galaxy. 

Given that GW190425 surprised us by being more massive than 
previously observed BNSs, it might also be possible that, despite 
our expectation, there is a class of BNS systems in which the individual NSs
have high spins. 
Until now, typical GW analyses of BNS systems are run with two different spin priors, 
one high-spin prior in which the NSs have dimensionless spins $|\chi| \lesssim 0.89$
and one "astrophysically-informed" low-spin prior in which the individual dimensionless 
NS spins are restricted by $|\chi| \lesssim 0.05$.
The latter is based on observations of pulsars in BNS systems, where 
the fastest-spinning BNSs, capable of merging within a Hubble time, are 
PSR J0737-3039A~\cite{Burgay:2003jj} and PSR J1946+2052~\cite{Stovall:2018ouw}.
Both will have dimensionless spins of $\chi \lesssim 0.04$ or $\chi \lesssim 0.05$ 
at merger, respectively. 
Contrary, the fastest-spinning neutron stars observed to date can have spins 
up to $\chi \sim 0.4$~\cite{Hessels:2006ze}, and the 
theoretical breakup spin for realistic EOSs is about $\chi\sim0.7$~\cite{Lo:2010bj}. 
Therefore, the fact that no millisecond pulsar has yet been observed 
in BNS systems, could just be an observational bias. 
While initially most numerical relativity (NR) studies have neglected spin effects, there has been a noticeable
advancement over the last few years in which different groups studied 
BNS configurations in which the individual NSs are spinning, see Refs.~\cite{Kastaun:2013mv,Bernuzzi:2013rza,Kastaun:2014fna,Dietrich:2015pxa,Tacik:2015tja,East:2015vix,
Dietrich:2016lyp,Kastaun:2016elu,Dietrich:2017xqb,Dietrich:2017aum,
Dietrich:2018upm,Dietrich:2018phi,Chaurasia:2018zhg,Most:2019pac,Ruiz:2019ezy,
Tsokaros:2019anx,East:2019lbk,Tichy:2019ouu}.
However, to our knowledge, none of these studies produced high-quality NR data for highly spinning systems, 
i.e., NR data with uncertainties small enough to validate and potentially improve 
existing GW models. 
However, it would be essential to perform such NR vs.\ GW model comparisons 
to ensure that the existing GW models can reliably describe also systems in 
larger regions of the BNS parameter space. 
Overall, a reliable analysis of detected GW signals relies on an 
accurate theoretical description to cross-correlates measured 
GW strain data with GW approximants~\cite{Veitch:2014wba} 
throughout the entire parameter space.
If employed waveform models were inaccurate, this would immediately lead to 
a systematic bias in the extraction of information from 
the observed data~\cite{Dudi:2018jzn,Samajdar:2018dcx,Samajdar:2019ulq,Gamba:2020ljo}. 

In this article we describe a first set of highly spinning, high-accuracy 
NR data of BNS systems. For this purpose, we simulate four different configurations. 
Details about the numerical setup and the employed configurations are presented 
in Sec.~\ref{sec:method}. In Sec.~\ref{sec:dynamics} we present a 
qualitative discussion of the merger dynamics and extract information about the 
ejecta and remnant properties. In Sec.~\ref{sec:analysis} we study the 
accuracy of our NR data and compare the extracted GW signals 
to a set of state-of-the-art GW models which are 
currently employed for the analysis of GW data. We finalize this study through 
an injection study to understand possible systematic biases.
We conclude in Sec.~\ref{sec:conclusion}. 
Unless otherwise stated, we employ geometric
units for which $c=G=M_\odot=1$.

%TTTTTTTTTTTTTTTTTTTTTTTTTTTTTTTTTTTTTTTTTTTTTTTTTTTTTTTTTTTTTTTTTTTTTTTT
\begin{table*}[t]
\caption{BNS configurations.
The first column gives the configuration name.
The next 6 columns provide the physical properties of the individual stars: employed EOS, the gravitational masses of the individual stars $M^{A,B}$, the
baryonic masses of the individual stars $M _b ^{A,B}$, 
the stars' dimensionless spin magnitudes $\chi^{A,B}$, 
and tidal deformabilities $\Lambda^{A, B}$. 
The remaining columns give the mass-weighted effective spin $\chi _\text{eff}$, 
the residual eccentricity $e$, the initial GW frequency $M\omega^0 _{22}$, the Arnowitt-Deser-Misner (ADM) mass $M_\text{ADM}$, and the angular
momentum $J$. 
The configurations were all evolved with the four 
resolution $n_{96}, n_{144}, n_{192}, n_{256}$.}
\label{tab:config}
\begin{tabular}{c|cccccc|cccccccc}
\toprule
Name & EOS &$M^{A,B}$ &  $M^{A, B}_b$    & $\chi^{A}$ & $\chi^B$ & $\Lambda^{A,B}$ & $\chi_{\rm eff}$  & $e$ & $M\omega ^0 _{22}$ &$M_\text{ADM}$ & $J$ \\
\hline
\Slyuuh & SLy  & $1.367$ & $1.495$ & $0.5759$ & $0.5759$ & $360.1$ & $0.5759$ & $0.0011$   & $0.038$  & $2.711$  &  $9.849$ \\
\Slyuu & SLy  & $ 1.357$  & $1.495$ & $0.3683$ & $0.3683$ & $376.7$  & $0.3683$ &  $0.00095$ &  $0.032$& $2.694$ & $9.343$\\
\Slyud & SLy &  $1.351$ & $1.495$ &  $0.1556$ & $-0.1556$ & $387.7$  & $0.0000$ & $0.00031$   &  $0.032$ & $2.682$& $8.036$ \\
\Slydd & SLy &  $1.354$ & $1.495$ & $-0.2775$ & $-0.2775$ & $382.7$  & $-0.2775$  & $0.00034$  & $0.032$   & $2.688$ & $7.148$\\
 \hline \hline
\end{tabular}
\end{table*}
%TTTTTTTTTTTTTTTTTTTTTTTTTTTTTTTTTTTTTTTTTTTTTTTTTTTTTTTTTTTTTTTTTTTTTTTT
%%%%%%%%%%%%%%%%%%%%%%%%%%%%%%%%%%%%%%%%%%%%%%%%
\section{Methods and Configurations}
\label{sec:method}
%%%%%%%%%%%%%%%%%%%%%%%%%%%%%%%%%%%%%%%%%%%%%%%%

\subsection{Physical Configurations}
\label{sec:method_config}
We study four different system with similar initial baryonic mass 
$M_b =1.495 M_\odot$, but different spin configurations. 
In the \Slyuuh{} configuration, each star has a spin of 
about $\chi^{\rm A, B} = 0.57$ aligned with the orbital angular momentum. 
In the \Slyuu{} configuration each star has aligned spin of about 
$\chi^{A,B} = 0.37$. 
In \Slyud{} one star has aligned spin and the other anti-aligned spin with 
magnitude $\chi^{A,B} = 0.155$ and \Slydd{} has anti-aligned spins of 
$\chi^{A,B}=-0.277$ for both stars. 
Further details are given in Tab.~\ref{tab:config}.
Spin effects are mostly characterized by the mass-weighted spin combination,
$\chi_{\rm eff} = (M^{A} \chi^{A} + M^{B} \chi^{B})/M$, 
where $\chi^i = |\vec{S}_i|/M^{i2}$ is the dimensionless spin parameter 
and $\vec{S}_i$ is the spin angular momentum of the $i^{th}$ NS.

\subsection{Numerical Setup}
\label{sec:method_setup}

Configurations simulated within this article, listed in 
Tab.~\ref{tab:config}, employ initial data constructed 
with the updated \texttt{SGRID} code~\cite{Tichy:2019ouu}. 
\texttt{SGRID}~\cite{Tichy:2009yr,Tichy:2011gw,Tichy:2012rp,Dietrich:2015pxa,Tichy:2016vmv,Tichy:2019ouu} 
uses surface fitting coordinates and solves the
conformal thin sandwich equations along with the constant 
rotational velocity approach to describe the NSs with 
arbitrary rotational profile. 
\texttt{SGRID} employs pseudospectral methods 
to solve the elliptic equations and the computational domain is divided into 38 patches 
(Fig.~2 of \cite{Tichy:2019ouu}). 
Through \texttt{SGRID}'s most recent update~\cite{Tichy:2019ouu},
we can construct initial data for configurations at the edge of the physically allowed 
BNS parameter space, which includes high spin, 
high mass-ratios and high compactness setups. 
In this work, we focus on the impact of high spins. 
We apply an eccentricity reduction procedure as described in 
Appendix B of Ref.~\cite{Tichy:2019ouu} to all our configurations to achieve 
target residual eccentricities below $\leq 10^{-3}$. 
The exact values are listed in Tab.~\ref{tab:config} along with the initial 
parameters such as initial ADM mass, 
angular momentum of the system, and initial GW frequency. The first and last lines in this table represent the highest aligned and anti-aligned spins that we were able
to obtain for the SLy EOS at the time with \texttt{SGRID}. As one can see,
achieving high anti-aligned spins is harder than high aligned spins.

\begin{figure}[t]
\includegraphics[width=0.5\textwidth]{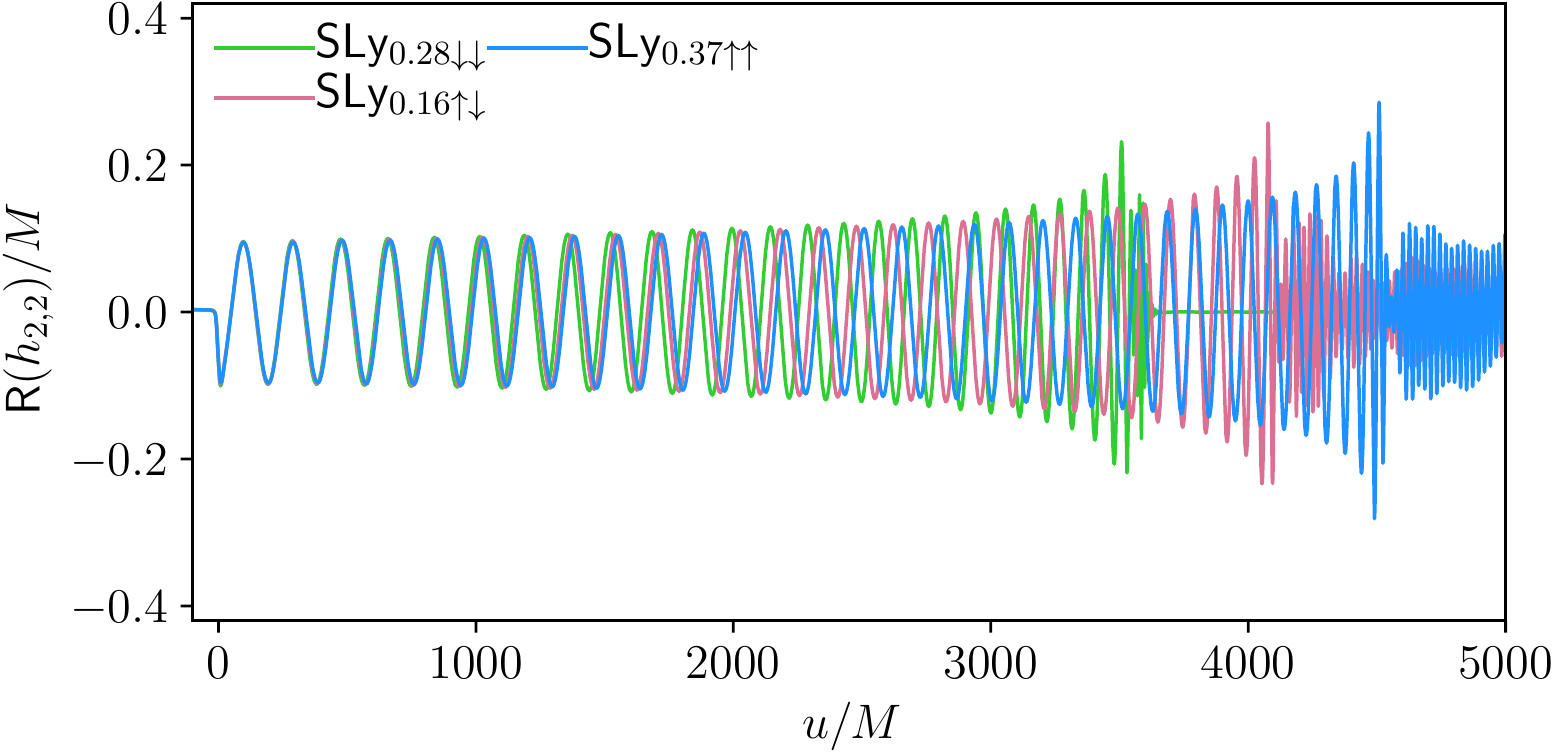}
\caption{Comparison of the highest resolution simulation for 
\Slyuu{}, \Slyud{}, and \Slydd{}. All these systems start at the same initial frequency, 
\Slyuuh{} is not shown because of a smaller initial frequency. 
Clearly visible is the orbital hangup effect caused by the interaction between the intrinsic spin 
of the NSs and the orbital angular momentum.}
\label{fig:GW_all}
\end{figure}

We evolve the initial data using the \texttt{BAM} code~\cite{Brugmann:2008zz,Thierfelder:2011yi,Dietrich:2015iva,Bernuzzi:2016pie}. 
and employ the Z4c formulation of the Einstein equation~\cite{Ruiz:2010qj,Hilditch:2012fp,Bernuzzi:2009ex} 
along with 1+log and gamma driver conditions~\cite{Bona:1994dr,Alcubierre:2002kk,vanMeter:2006vi} 
for the evolution of the lapse and shift vector. 
For the evolution of the matter variables, we use a $3+1$ conservative 
Eulerian formulation of general-relativistic hydrodynamics by 
defining Eulerian conservative variables from the rest-mass 
density $\rho$, pressure $p$, internal energy $\epsilon$, and 3-velocity $v^i$.
The system is closed using a piecewise-polytropic fit for the 
SLy~\cite{Read:2008iy} EOS with an additional thermal contribution 
to the pressure given by $p_{th} = (\Gamma_{th} -1)\rho\epsilon$, 
where we set $\Gamma_{th} = 1.75$ ~\cite{Bauswein:2010dn,Zwerger:1997ret}. We note that SLy is a rather soft EOS, 
i.e., it supports NSs with tidal deformabilities 
$\tilde{\Lambda} \approx 400$, for a $1.35 M_\odot$ mass NS
and is in agreement with current observations~\cite{Margalit:2017dij,Bauswein:2017vtn,Radice:2017lry,Rezzolla:2017aly,
Most:2018hfd,Ruiz:2017due,Radice:2018ozg,Hinderer:2018pei,Capano:2019eae,Raaijmakers:2019qny,
Breschi:2021tbm,Coughlin:2018miv,Coughlin:2018fis,Dietrich:2020efo,Huth:2021bsp}.

Our numerical domain is divided into a hierarchy of cell centered nested Cartesian grids 
consisting of $L$ levels labeled by $l = 0,..., L-1$. 
Each level $l$ contains one or more Cartesian boxes 
with constant grid spacing $h_l$ and $n$ (or $n_{mv}$) number of points per direction. 
The resolution in each level is given as $h_l = h_0 /2^l$. 
Levels $ l \geq l_{mv}$ can move dynamically according to the technique of 'moving boxes'; 
here we employ $l_{mv} = 5$. 

The BAM grid setup considered in this work consists of seven refinement levels. 
We use four different resolutions for each configuration labeled 
$n_{96}$, $n_{144}$, $n_{192}$, and $n_{240}$, 
where the subscript refers to the number of points 
in the finest refinement box covering the NSs. 
This leads to a finest grid spacing for \Slyuuh{} and \Slyud{} of 0.075 $M_{\odot}$ 
with respect to the highest resolution $n_{240}$.
For \Slyuu{} and \Slydd, the finest grid spacings 
are 0.068 $M_{\odot}$ for the highest resolution. 
All setups start at an initial frequency of $M\omega = 0.032$, 
except for \Slyuuh{} with an initial frequency of $M\omega = 0.038$. 
This difference is caused by the fact that we `reuse' the initial data for \Slyuuh{} 
computed in Ref.~\cite{Tichy:2019ouu}.

%%%%%%%%%%%%%%%%%%%%%%%%%%%%%%%%%%%%%%%%%%%%%%%%%%%%%%%%%%%%%%%%%%%%%%%%%%
\section{Merger Dynamics}
\label{sec:dynamics}
%%%%%%%%%%%%%%%%%%%%%%%%%%%%%%%%%%%%%%%%%%%%%%%%%%%%%%%%%%%%%%%%%%%%%%%%%%
For the analysis of the binary evolution, in particular the study of 
ejecta and disk mass estimation, and the emitted GWs
signal, we use methods described in detail in Refs.~\cite{dbt_mods_00029293,Dietrich:2016hky,Dietrich:2016lyp,
Chaurasia:2018zhg,Chaurasia:2020ntk}. 

\subsection{Qualitative Discussion}
\label{sec:qualitative}

\begin{figure*}[t]
\includegraphics[width=0.8\textwidth]{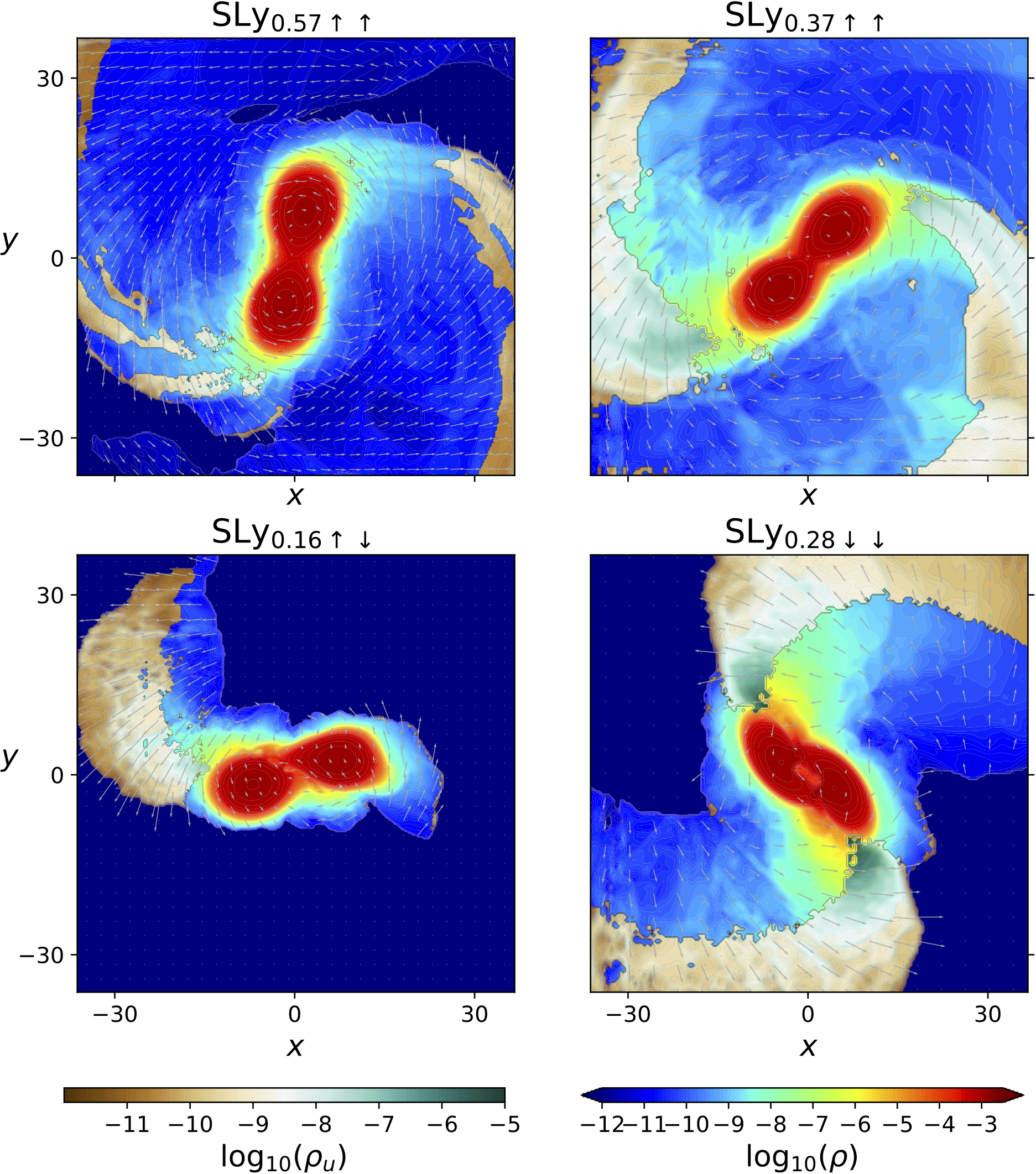}
\caption{Rest-mass density profile and velocity field inside the orbital plane for all simulations. The snapshots represent moments close to the merger. The rest-mass density $\rho$ is shown on a logarithmic scale from blue to red. The rest-mass density of unbound material ($\rho_u$) is colored from brown to dark green. Most material gets ejected from the tidal tails of the NSs inside the orbital plane.}
\label{fig:2d_rho}
\end{figure*}

Although spin effects that are present during the BNS coalescence 
have been studied before, cf.~\cite{Kastaun:2013mv,Bernuzzi:2013rza,Kastaun:2014fna,Dietrich:2015pxa,Tacik:2015tja,East:2015vix,
Dietrich:2016lyp,Kastaun:2016elu,Dietrich:2017xqb,Dietrich:2017aum,
Dietrich:2018upm,Dietrich:2018phi,Chaurasia:2018zhg,Most:2019pac,Ruiz:2019ezy,
Tsokaros:2019anx,East:2019lbk,Tichy:2019ouu}, 
we want to briefly summarize some of the main features that 
dominate the overall dynamics. 

As visible in Fig.~\ref{fig:GW_all}, we see that systems with aligned spin 
show a delayed merger, while systems with anti-aligned spin show an earlier merger. 
This so called hang-up effect of spin-aligned systems is caused by the interaction 
between the orbital angular moment and the intrinsic spin of the NSs; cf.~\cite{Blanchet:2013haa} and references therein. 

Interestingly, we see in the 2d-density plots shown in Fig.~\ref{fig:2d_rho} 
a noticeable difference at the time of merger. Most notably, 
the anti-aligned configuration shows a clear density minimum in the center. 
Such a minimum is less pronounced in all other evolved configurations. 
Furthermore, the shape of the individual NS is highly deformed. 
This deformation, while in the shown figure being also coordinate dependent, 
hints towards an enhancement of non-equilibrium tides. 
This observation suggests that due to the large anti-aligned spin
spin-dependent dynamical tidal effects
(as discussed in~\cite{Steinhoff:2021dsn}) have to be included 
for an accurate modeling of the system; cf.~Sec.~\ref{sec:comparison}.  
Indeed, tidal effects become dynamically enhanced close to a resonance
between the orbital motion and the fundamental oscillation mode of the star,
which is most prominent for an anti-aligned spin configuration since the
resonance occurs at lower frequency in this case. This theoretical expectation
together with the density minimum at the center make it plausible that the
visible strong deformation of the stars is not a mere coordinate effect.

%%%%%%%%%%%%%%%%%%%%%%%%%%%%%%%%%%%%%%%%%##
\subsection{Ejecta and Remnant properties}
\label{sec:ejecta}
%%%%%%%%%%%%%%%%%%%%%%%%%%%%%%%%%%%%%%%%%##

In addition to the bound rest-mass density,
Fig.~\ref{fig:2d_rho} shows the unbound rest-mass density (ejecta) for each simulation.
One finds that most of the mass ejection originates from the tidal tail behind the stars, and only a small amount of mass is ejected through shocks. 
%such ejecta is mostly shock driven and happens in a perpendicular 
%direction to the orbital plane. 
Overall, we observe that for aligned spin configurations more
material gets ejected than for anti-aligned systems. 

\begin{figure}
\includegraphics[width=0.4\textwidth]{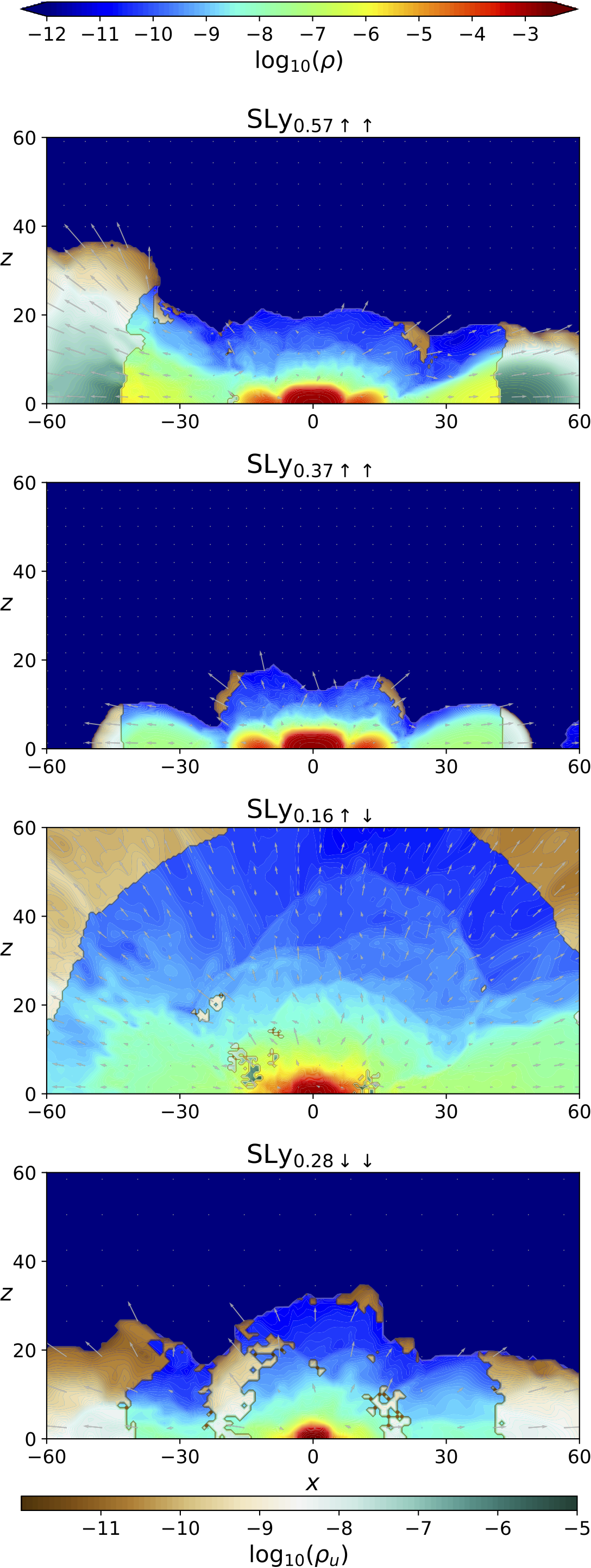}
\caption{Bound and unbound rest-mass density and velocity profile in the x-z-plane. 
The snapshots are taken $10\rm ms$ after the merger. The rest-mass density $\rho$ is shown on a logarithmic scale from blue to red. The rest-mass density of unbound material ($\rho_u$) is colored from brown to dark green. Most material gets ejected from the tidal tails of the NSs inside the orbital plane, but in case of anti-aligned spin, some material is also ejected orthogonal to the orbtital plane.}
\label{fig:2d_rho_xz}
\end{figure}

The ejecta mass computation uses two different methods; see~\cite{Chaurasia:2018zhg} for more details. 
The first method is based on a volume integration of the unbound matter, 
$M_{\rm ej}^{\nu}$, where the unbound matter is defined through the two conditions
\begin{equation}
    u_t < -1 \quad \text{and} \quad v^ix_i > 0,
\end{equation}
where $u_t = -W(\alpha - \beta_iv^i)$ is the time component of the fluid 4-velocity, 
$\alpha$ is the lapse, $\beta^i$ is the shift vector, $W$ is the Lorentz factor, and $x^i = (x, y, z)$ 
the coordinate vector.
The second method uses the matter flux across a coordinate sphere with radius $r_s$,
\begin{equation}
    M_{\rm ej}^S = \int_0^t dt' \int_{r=r_s} [D_u(\alpha v^i + \beta^i)n_i]r^2 d\Omega,
\end{equation}
with $n_i = x_i/r$ and $r = \sqrt{x^ix_i}$. 
$D_u$ denotes the unbound fraction of conserved rest mass density $D=W \rho$. 

The estimated ejecta mass is listed in Tab.~\ref{tab:ejecta} for all configurations. 
We find that the highest ejecta mass of about $5\times10^{-2}M_\odot$ 
(independent of the resolution) is present for the \Slyuuh{} configuration. 
This observation is in agreement with, e.g,~\cite{Dietrich:2016lyp,Kastaun:2016elu,Ruiz:2019ezy}, 
while~\cite{East:2019lbk,Most:2019pac} point out that in some cases anti-aligned 
setups can eject more massive ejecta. 
This `disagreement' is likely caused by the observation that different ejecta mechanisms have different spin dependence. 
While Ref.~\cite{Dietrich:2016lyp} explained that aligned-spin systems 
will create larger dynamical ejecta due to larger torque in the tidal tails,
Ref.~\cite{Most:2019pac} points out that anti-aligned spin can lead 
to a larger impact velocity around the moment of merger, 
which increases shock-driven outflows.  
Both observations are in agreement with our simulations as visible in 
Fig.~\ref{fig:2d_rho_xz}, where we show the bound and unbound density 
$10\rm ms$ after the merger. One finds that setups with 
aligned spin produce signifnicantly more ejecta inside the orbital, 
but aligned spin setups, most notably \Slydd{}, produce ejecta orthogonal to the orbital plane. 

We note that furthermore, disk-wind ejecta released after the merger will also
depend on the spin of the NSs. Generally, the initially aligned 
spin systems create a faster rotating remnant, a more massive debris disk (see Fig.~\ref{fig:2d_rho_xz}), and larger ejecta mass. 
However, longer simulations which have to include
more advanced microphysical descriptions will be 
necessary for quantitative studies. 

\begin{table}[t]
\caption{Ejecta mass estimates using the volume integral 
$M_\text{ej}^{\mathcal{V}}$ and the coordinate sphere integration 
$M_{\rm ej}^S$ for the two highest resolutions $n_{240}$ and $n_{192}$.}
\begin{tabular}{c|cc|cc}
\toprule
\hline \hline
Name &  \multicolumn{2}{c}{$M_\text{ej}^{\mathcal{V}} (M_{\odot})$} &  \multicolumn{2}{c}{$M_\text{ej}^{S} (M_{\odot})$}  \\ 
           &  $n_{240}$  & $n_{192}$  & $n_{240}$ & $n_{192}$\\   

\hline

\Slyuuh    &  $0.0506$  & $0.0506$ & $0.0549$ &  $0.0559$\\ 
\Slyuu      & $0.0090$  & $0.0096$ & $0.0062$  & $0.0083$ \\ 
\Slyud     & $0.0026$ & $0.0028$ & $0.0053$  & $0.0022$ \\ 
\Slydd      & $0.0081$ & $0.0064$  &  $0.0003$  & $0.0007$ \\ 

\hline \hline
\end{tabular}
\label{tab:ejecta}
\end{table}

Another major difference between the spin-aligned and anti-aligned configurations is the 
stability of the formed remnant, see Refs.~\cite{Dietrich:2020eud,Bernuzzi:2020tgt}.
In fact, the anti-aligned spin configuration \Slydd{} is the only setup forming a black hole quickly after the merger. 
For this setup, the final black hole has a mass of about $M_{\rm BH} = 2.641 M_\odot$ 
and a dimensionless spin of $\chi_{\rm BH} = 0.713$. 
Except for \Slydd, all other configurations form hypermassive neutron stars, 
which do not collapse to a black hole until the end of our simulations. 
This difference in the collapse time is related to the reduced total 
angular momentum for the \Slydd{} configuration, so that due to the reduced
momentum support a black hole is quickly formed. 
A similar effect with respect to the remnant's lifetime has 
also been discussed in Refs.~\cite{Kastaun:2013mv,Bernuzzi:2013rza,Dietrich:2016lyp,Ruiz:2019ezy}, but due to the large spin contributions considered here, 
has not been so pronounced. 
Indeed, this suggests that remnant classifications and 
also classifications of the prompt collapse threshold~\cite{Bauswein:2013jpa,Agathos:2019sah,Koppel:2019pys,Bauswein:2020xlt} should contain the intrinsic spin 
of the individual NSs if they are employed within the entire parameter space.

\begin{figure*}[t]
\begin{tabular}{cc}
\subfloat[\Slyuuh]{\includegraphics[width=0.5\textwidth]{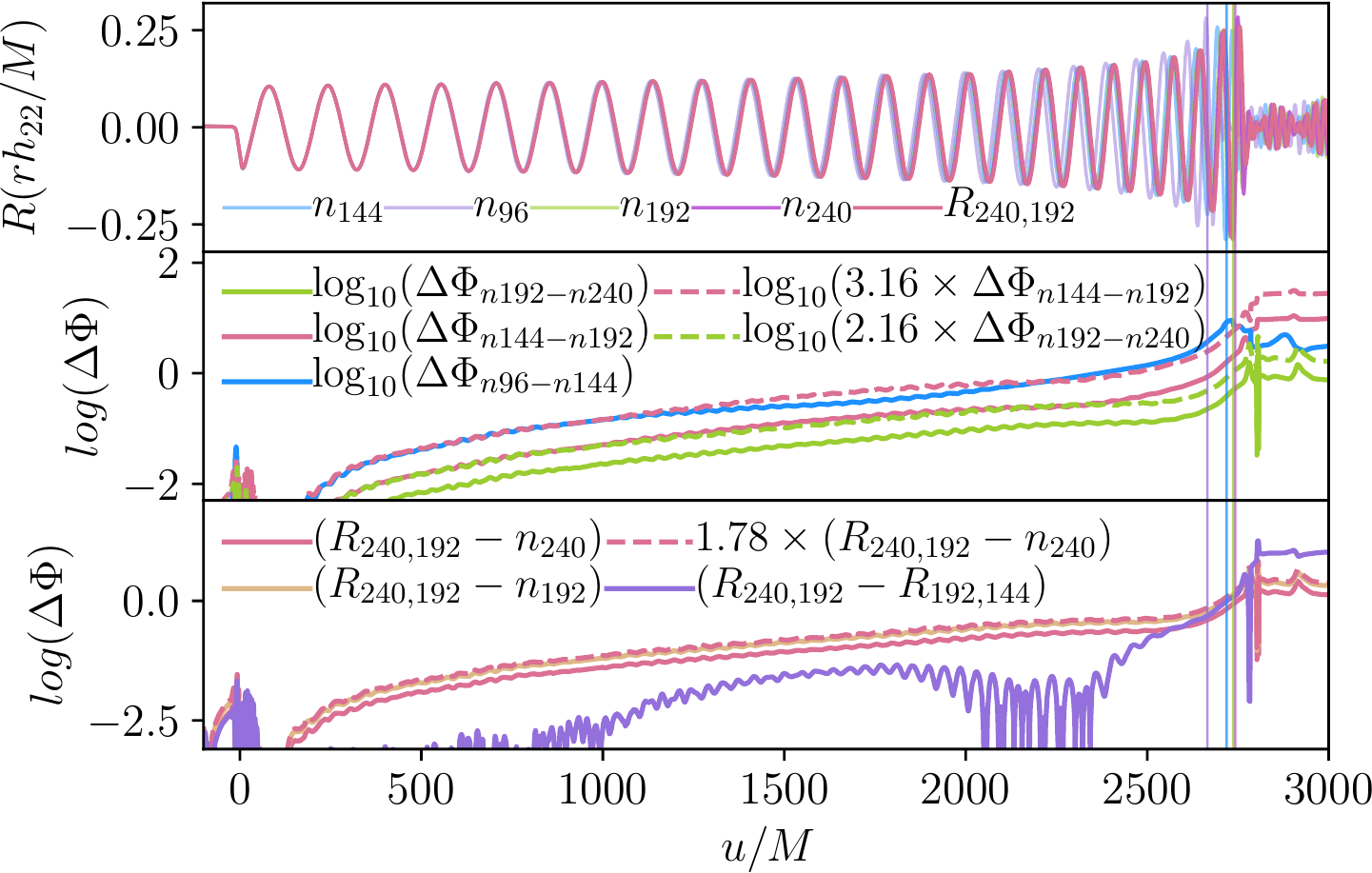}} & 
\subfloat[\Slyuu]{\includegraphics[width=0.5\textwidth]{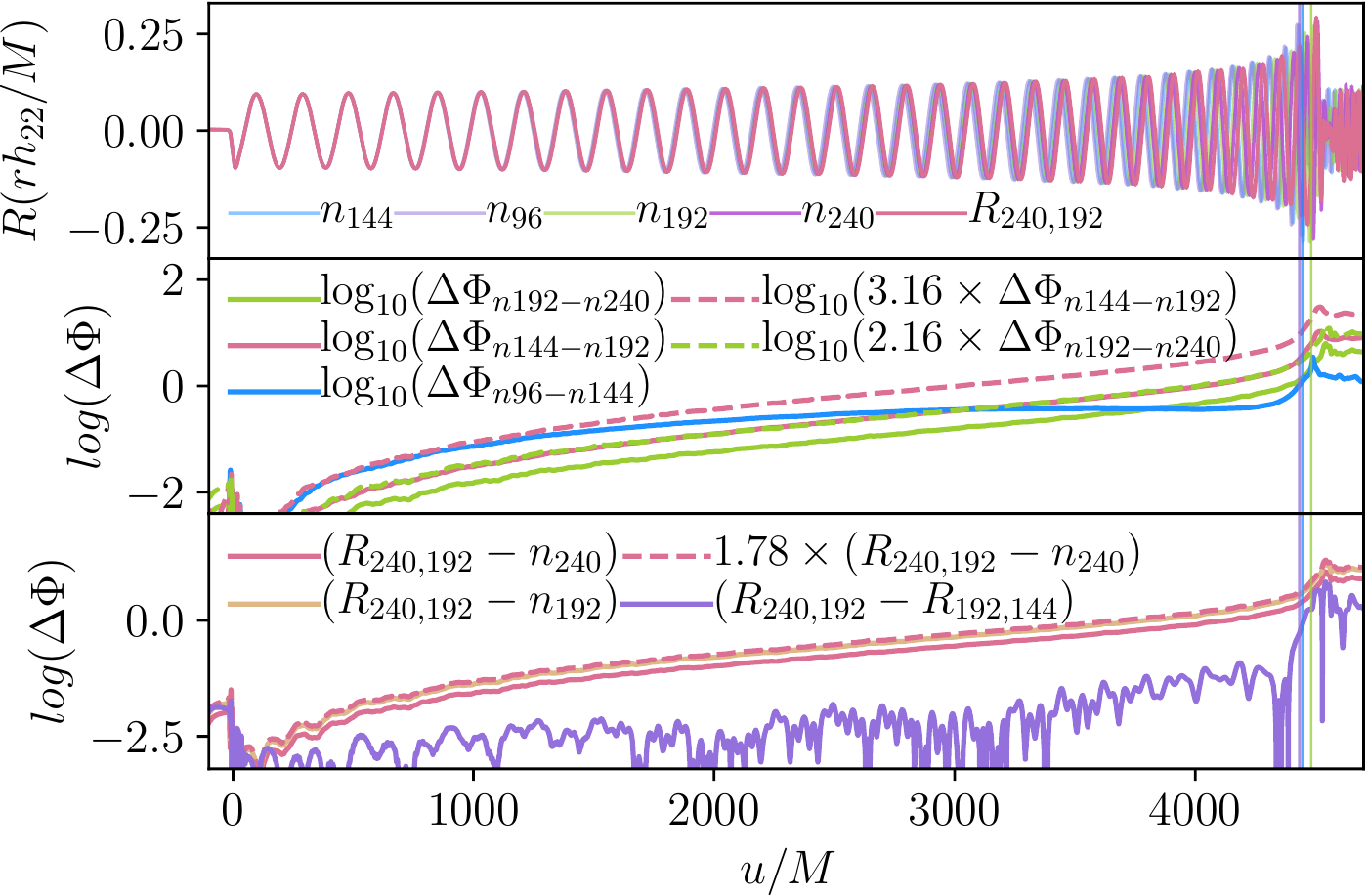}} \\
\subfloat[\Slyud]{\includegraphics[width=0.5\textwidth]{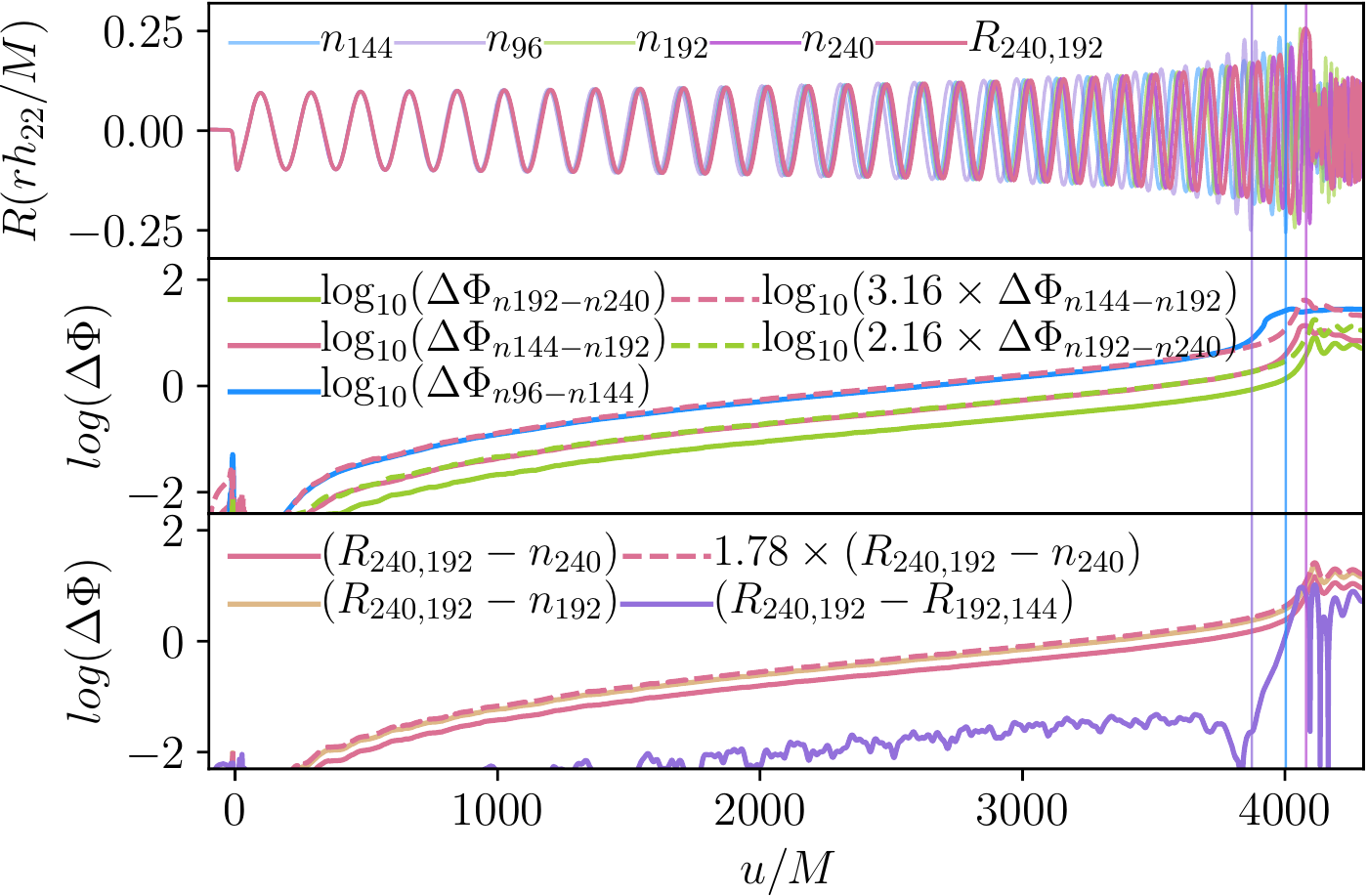}} & 
\subfloat[\Slydd]{\includegraphics[width=0.5\textwidth]{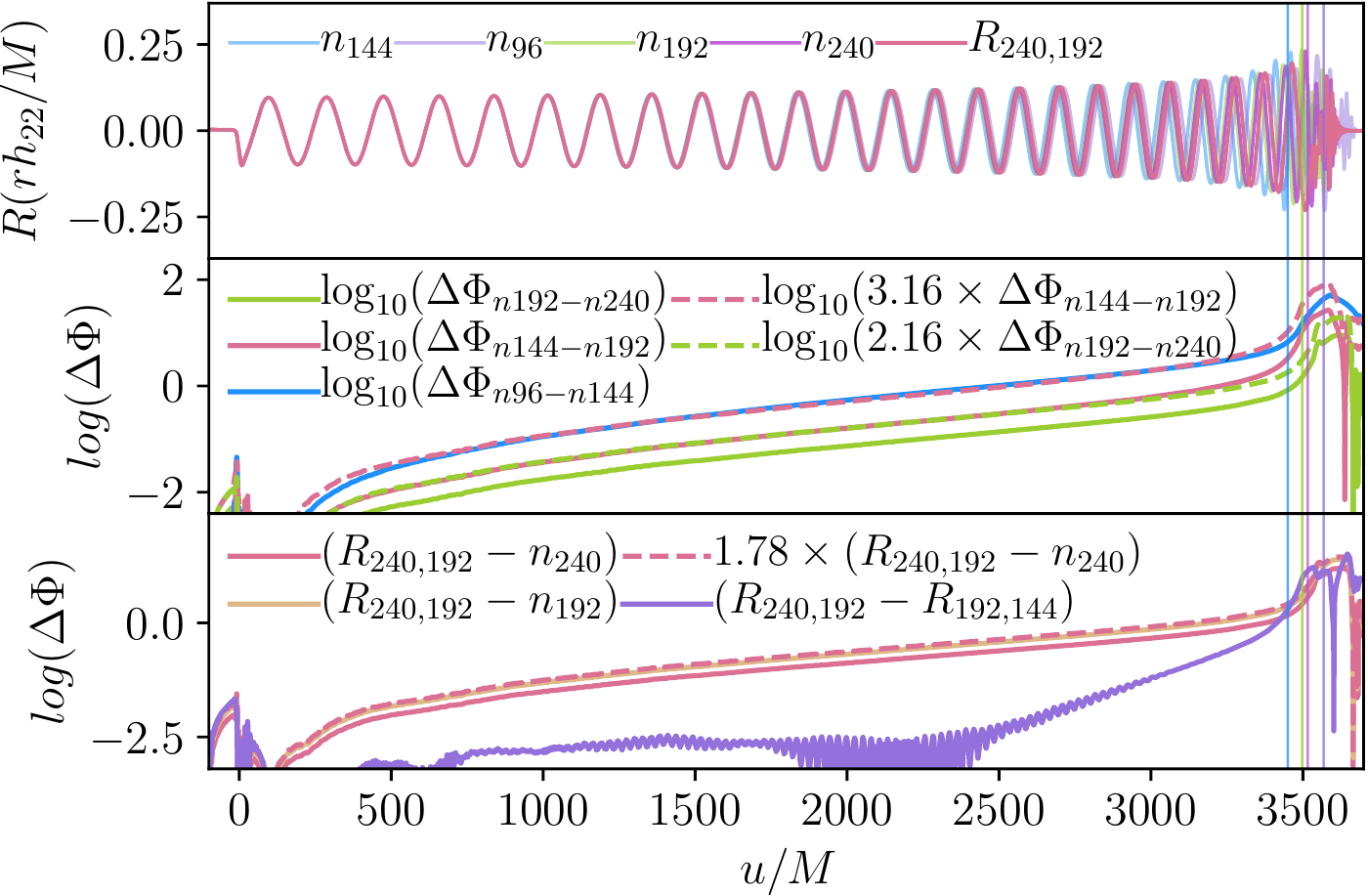}}\\
\end{tabular}
\caption{Top panels: Real part of the GW signal for the four different resolutions employing 96, 144, 192 and 240 points in the refinement levels covering the individual NSs for configurations \Slyuuh (top left panel), \Slyuu (top right panel), \Slyud (bottom left panel), and \Slydd (bottom right panel). Middle panels: Phase difference between different resolutions. Bottom panels: Phase difference between different Richardson-extrapolated waveforms or between a Richardson-extrapolated waveform and the waveform from an individual resolution. The vertical lines in each panel refer to the time of merger, i.e., the time of the maximal GW amplitude for the individual resolutions. The dashed lines in the bottom two panels show the phase difference scaled to the next lowest pair of resolutions assuming second order convergence. $u/M$ denotes the retarded time scaled by the total mass $M$.}
\label{fig:richardson}
\end{figure*}

%%%%%%%%%%%%%%%%%%%%%%%%%%%%%%%%%%%%%%%%%%%%%%%%%%%%%%%%%%%%%%%%%%%%%%%%%%
\section{Analysis and Model Comparison}
\label{sec:analysis}
%%%%%%%%%%%%%%%%%%%%%%%%%%%%%%%%%%%%%%%%%%%%%%%%%%%%%%%%%%%%%%%%%%%%%%%%%%

\subsection{Convergence of the GW signal}
\label{sec:convergence}

In Fig.~\ref{fig:richardson}, we test the convergence properties for all configurations 
listed in Tab.~\ref{tab:config}. 
The usage of multiple grid resolutions and setups, 
i.e., 96, 144, 192, and 240 points in the refinement levels 
covering each individual NS, allows us to test convergence properties
of the GW signal extracted from our simulation. 
As discussed in, e.g., Refs.~\cite{Bernuzzi:2016pie,Dietrich:2018upm} 
the full error budget of the GW signal would consist of a number of individual effects, 
most notably the finite extraction radius of the waves and the discretization 
error introduced through the finite resolution of the numerical domain. 
While uncertainties through finite extraction 
radii are typically of the order of 
$\lesssim 0.1 \rm rad$, finite resolution errors easily 
contribute on the order of a radian to the final error budget. 

However, if we are able to determine the exact 
convergence order of our simulation, we can employ
Richardson extrapolation to obtain an improved estimate 
for the GW signal, labeled as $R_{240, 192}$.
Based on this consideration, 
we test in Fig.~\ref{fig:richardson} the convergence of the GW phase 
and also present the phase difference between the highest resolution 
and the Richardson-extrapolated waveform. 
This phase difference can be understood as an estimate of 
the numerical uncertainty of the gravitational waveform, 
where we will neglect the influence of the finite radius extraction 
due to its smaller size. 

Regarding Fig.~\ref{fig:richardson}, subfigure (a) shows data 
corresponding to the \Slyuuh{}  configuration. 
For this (as well as for all other systems), 
the upper most parts show the $(2, 2)-$ mode of GWs for all 
employed resolutions and for the Richardson-extrapolated 
waveform $R_{240, 192}$. 
The vertical solid lines indicate the merger time. 
The middle panel shows the phase difference among 
different resolutions (solid lines), 
as well as the rescaled phase difference according 
to an assumed second order convergence (dashed lines). 
The bottom panel shows the phase difference between 
the Richardson-extrapolated data and two highest resolutions. 
Also for these phase differences, 
we find clear second order convergence, 
which simply acts as a cross check for the 
correctness of the Richardson extrapolation. 
The purple curve in the bottom panels shows 
the phase difference between two Richardson-extrapolated 
waveforms based on data from different NR resolutions. 
This phase difference shows no clear monotonic behavior 
and contains even zero crossings. 
This fact indicates that the numerical accuracy of our 
data would not allow us to perform a second 
Richardson-extrapolation step to further improve our 
final estimate. 
Similarly, we show in Fig.~\ref{fig:richardson}
similar plots for \Slyuu{} in the top right panel 
(b), for \Slyud{} in the bottom left panel 
(c), and for \Slydd{} in the bottom right panel (d) of Fig.~\ref{fig:richardson}. 
Second order convergence is obtained throughout the 
inspiral for all the configurations when we consider the highest NR resolutions.

The final difference between the highest 
resolution and the Richardson-extrapolated waveform is 
then for all cases used to estimate the NR uncertainty of the simulation.

\subsection{GW Model Comparison}
\label{sec:comparison}

\begin{figure}[t]
\includegraphics[width=0.5\textwidth]{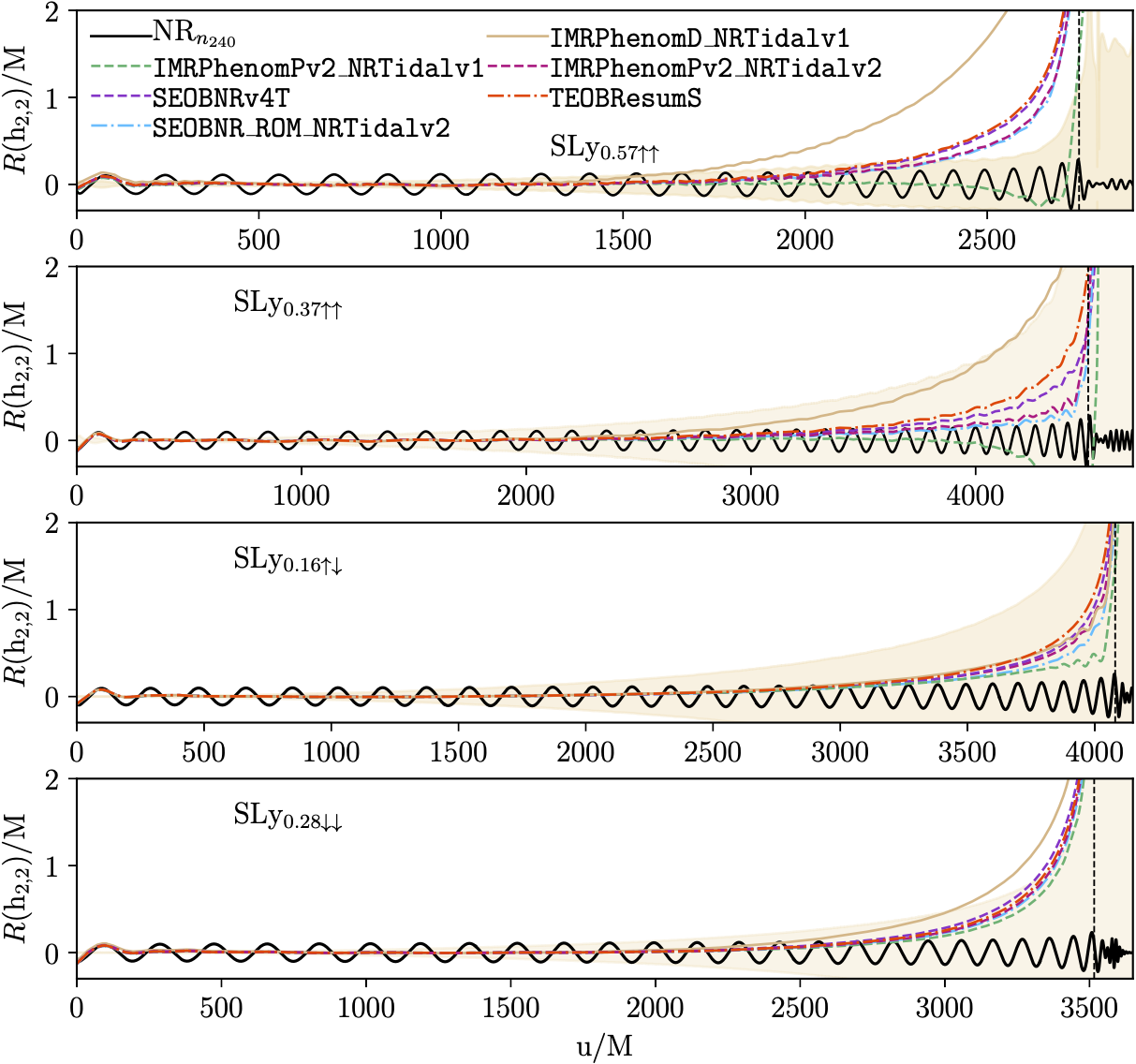}
\caption{We plot the highest NR resolution waveform of the GW signal for the four different configurations \Slyuuh (top left panel), \Slyuu (second top panel), \Slyud (third from the top panel) and \Slydd (bottom panel) with black curve. For each configuration, we compare the GW waveform with different waveform model such as \IMRPhenomPNRtidal2 , \SEOBT , \SEOBNRROMNRtidal , \IMRPhenomP1NRtidal ,  \TEOBResumS , and \IMRPhenomDNRtidal , respectively. The  phase difference between different highest NR resolution waveforms and waveform models is given by different colors. The alignment window is between $u/M=[200, 1500]$ for all configurations. The vertical black dashed line in each panel refers to the time of merger, i.e., the peak time of the GW amplitude. $u/M$ denotes the retarded time scaled by the total mass $M$.}
\label{fig:compareGW}
\end{figure}

\begin{figure}[t]
\includegraphics[width=0.5\textwidth]{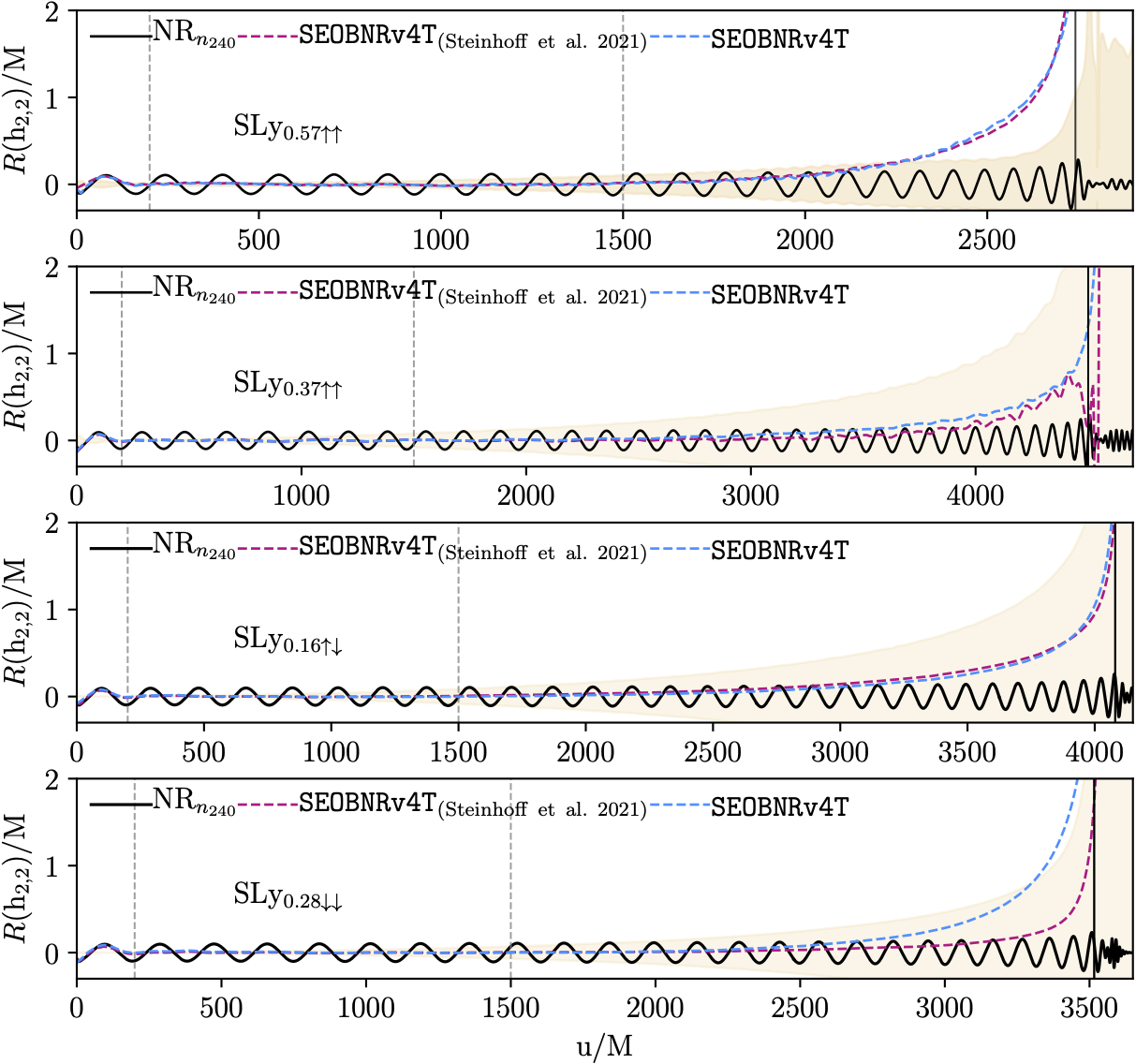}
\caption{Phase difference for all our setups employing the original \SEOBT and the updated \SEOBT model as developed in~\cite{Steinhoff:2021dsn}.}
\label{fig:SEOBNRTnew}
\end{figure}

Since the NR dataset produced for this article contains BNS simulations with high individual NS spins, uses low eccentricity (below $10^{-3}$) initial data, and since we have the possibility for a clear error assessment, it is natural to use this dataset for waveform model comparison. We will compare our numerical data with a set of waveform models implemented in the LALSuite package~\cite{lalsuite}, and we will summarize the main features of the model in the following.
  
\IMRPhenomDNRtidal: \IMRPhenomD is a phenomenological,
frequency-domain waveform model discussed in detail in
Refs.~\cite{Husa:2015iqa,Khan:2015jqa}. It describes non-precessing
BBH coalescences throughout inspiral, merger, and ringdown.
To obtain BNS waveforms, the \IMRPhenomD  approximant is augmented with tidal phase corrections given by \NRTidal , Refs.~\cite{Dietrich:2017aum,Dietrich:2018uni}.
No additional contributions from EOS-dependent spin-spin or cubic-in-spin effects are present, i.e., within this model the quadrupole moment is set to 1.

\IMRPhenomP1NRtidal: This model is based on \texttt{IMRPhenomPv2}, which describes precessing \ac{BBH} systems throughout the inspiral, merger and ringdown~\cite{Hannam:2013oca}. \IMRPhenomPv2 is augmented with the \NRTidal phase corrections to model BNS mergers~\cite{Dietrich:2017aum,Dietrich:2018uni}. 
In addition, the model includes 2PN and 3PN spin-spin corrections that depend on 
the EOS-dependent spin-induced quadrupole moment. 

\IMRPhenomPNRtidal2: The updated version of \IMRPhenomP1NRtidal model is \IMRPhenomPNRtidal2. It uses \NRtidal2 tidal phase corrections which include up to 3 PN spin-spin effects, including quadrupole and octupole contributions up to 3.5 PN as well as cubic-in spin effects and a tidal amplitude correction.

\SEOBNRROMNRtidal: This approximant is based on an EOB description of the general-relativistic two-body problem ~\cite{Buonanno:1998gg, Buonanno:2000ef} with free coefficients tuned to NR waveforms ~\cite{Bohe:2016gbl}. The BBH model \texttt{SEOBNRv4\_ROM} is augmented with the \NRtidal2 phase corrections ~\cite{Dietrich:2019nrt} to obtain BNS waveforms. It also includes spin-corrections and tidal amplitude corrections similar to \IMRPhenomPNRtidal2.

\SEOBT: This model is a time domain EOB model~\cite{Hinderer:2016eia,Steinhoff:2016rfi} which includes the quadrupolar and octopolar adiabatic and dynamical tides, spin-induced quadrupole moment effect, with a prescription for tapering at the end of the waveform. 
In this paper, we include two different versions of the model. First, the currently implemented one in the LALSuite, second an updated version that was recently presented in~\cite{Steinhoff:2021dsn} and incorporates spin-dependent 
resonance effects that change the dynamical tides description. 

\TEOBResumS: \TEOBResumS incorporates an enhanced
attractive tidal potential derived from resummed PN and gravitational self-force expressions of the EOB $A$-potential that determined tidal interactions Ref.~\cite{Nagar:2018zoe, Bernuzzi:2014owa, Damour:2009wj}.
For BNS, tidal effects are incorporated by computing a resummed attractive potential such that the tidal phase includes next-to-leading order (NLO) tidal contributions and gravitational self-force description of relativistic tidal interactions. It incorporates the EOS dependent self-spin effects up to NLO.\\

In Fig.~\ref{fig:compareGW}, we compare our NR data with all previously described waveform models. 
For this purpose, we show the estimated uncertainty of the NR dataset as shaded regions and the phase difference between the NR data and the GW approximants as dashed lines. In general, we find that all models tend to underestimate tidal effects with respect to the NR data.
This phenomenon was already discussed and outlined in a number of works, e.g., \cite{Baiotti:2010xh,Bernuzzi:2014owa,Dietrich:2017feu,Akcay:2018yyh} and has two reasons. 
On the one hand, analytical models generally underestimate tidal contributions during the last stages of the coalescence due to missing higher order corrections and physical effects appearing when both stars come into contact. On the other hand, NR simulations tend to overestimate tidal effects because numerical dissipation leads to an accelerated inspiral, which emphasizes the importance of a confident error estimate.

The top panel of Fig.~\ref{fig:compareGW} shows the comparison against \Slyuuh. 
We find that \IMRPhenomDNRtidal has the largest dephasing with respect to the NR
data. This observation is not surprising since this model does not incorporate EOS-dependent spin-spin interactions. These interactions appear at first at 2PN order and, for spin magnitudes  employed in this work, can lead to large dephasings. Contrary \IMRPhenomP1NRtidal shows the smallest dephasing with respect to the NR setup, where we expect that this is caused (i) by the fact that the original NRTidal contribution is more attractive than the updated NRTidalv2 contribution and (ii) that, in contrast to the NRTidalv2 approximants no 3.5PN spin-spin is included. Both contributions lead to an accelerated inspiral and hence a better agreement with the NR data. 

The second panel shows a comparison against \Slyuu. 
As before, \IMRPhenomDNRtidal shows the largest dephasing and is outside the NR uncertainty. 
The other models stay within the estimated NR uncertainty up to the moment of merger. 

For the \Slyud{} setup (third panel of Fig.~\ref{fig:compareGW}) 
all models stay within the estimated NR uncertainty. 
This indicates that for systems with $\chi_{\rm eff} \approx 0$, 
but still large individual spin contributions, 
the existing BNS waveform models might allow a reasonable description of the last few orbits covered by our NR simulations. 

Finally, results for \Slydd{} are shown in the bottom panel. 
It becomes clear that none of the existing models that is used for the GW analysis of 
previous events is able to describe all systems reliably.
%We find that about 7 GW cycles before the merger, 
%the dephasing is already larger than the estimated NR uncertainty. 

Interestingly, the recent model update of \SEOBT~\cite{Steinhoff:2021dsn} seems to be capable of 
describing \Slydd{}, i.e., systems with large anti-aligned spin contribution very well. 
We show a comparison of this setup and others in Fig.~\ref{fig:SEOBNRTnew}, where the model which includes
a spin-dependent description of the tidal section through the 
inclusion of resonance effects is clearly advantageous in the anti-aligned case.

\begin{figure*}[t]
\includegraphics[width=0.9\textwidth]{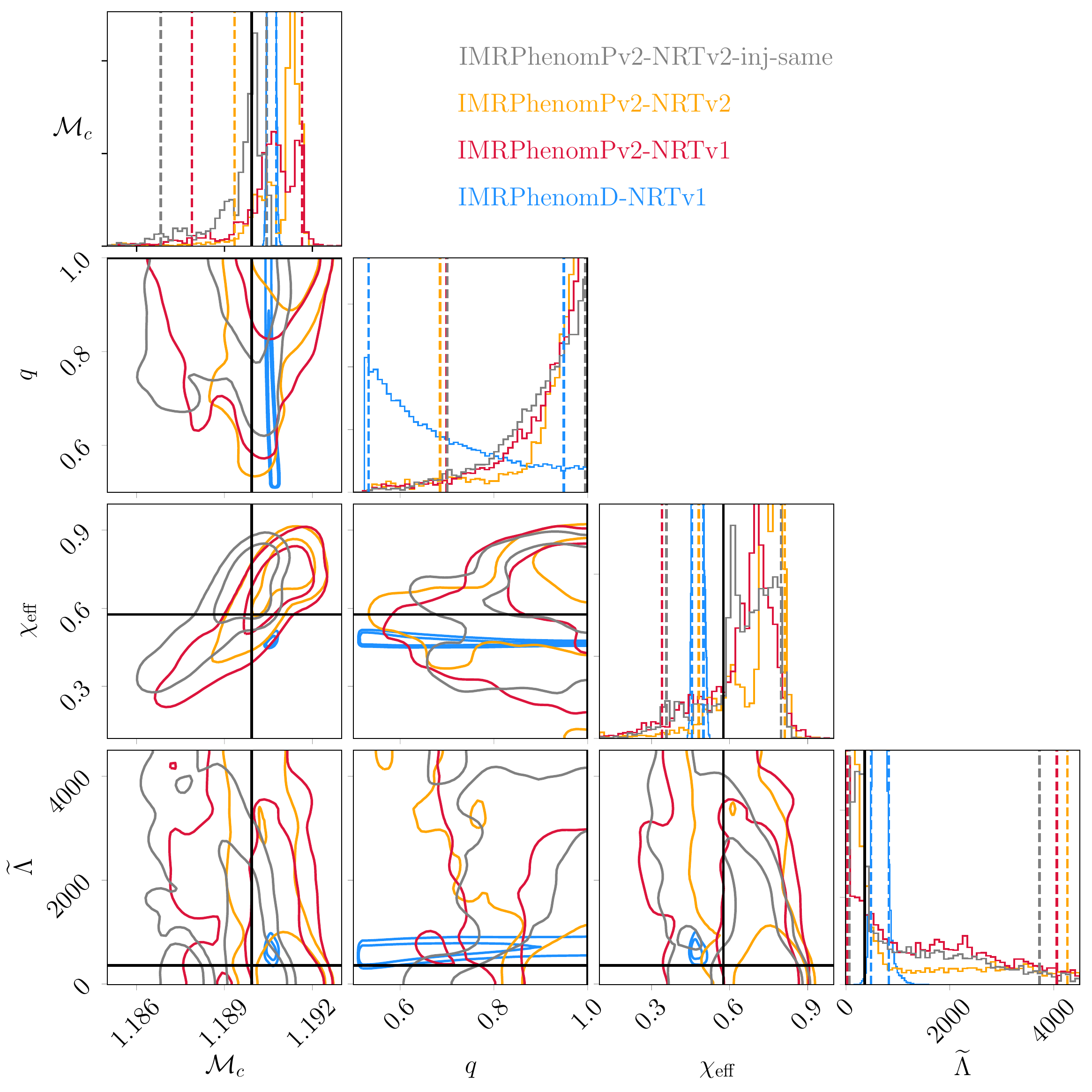}
\caption{Estimated chirp mass, mass ratio, effective spin, and tidal deformability for the 
         \Slyuuh{} setup. Recoveries labeled by \IMRPhenomP1NRtidal (red), \IMRPhenomPNRtidal2 (orange), \IMRPhenomDNRtidal (blue) use different approximants to recover our hybridized waveform which is a combination of the updated \SEOBT model and the highest NR resolution data. The \IMRPhenomPNRtidal2-inj-same (gray) uses for injection and recovery the \IMRPhenomPNRtidal2 model with the same source parameters as the hybrid. The latter injection serves as validation set for our injection setup. 
         The injected source parameters are marked with a solid black line. 
         The contours shown in the 2D-plots refer to 50\% and 90\% credible intervals, 
         intervals marked in the 1D plots refer to 90\% credible intervals.}
\label{fig:PE_sly5uu}
\end{figure*}

\begin{figure*}[t]
\includegraphics[width=0.9\textwidth]{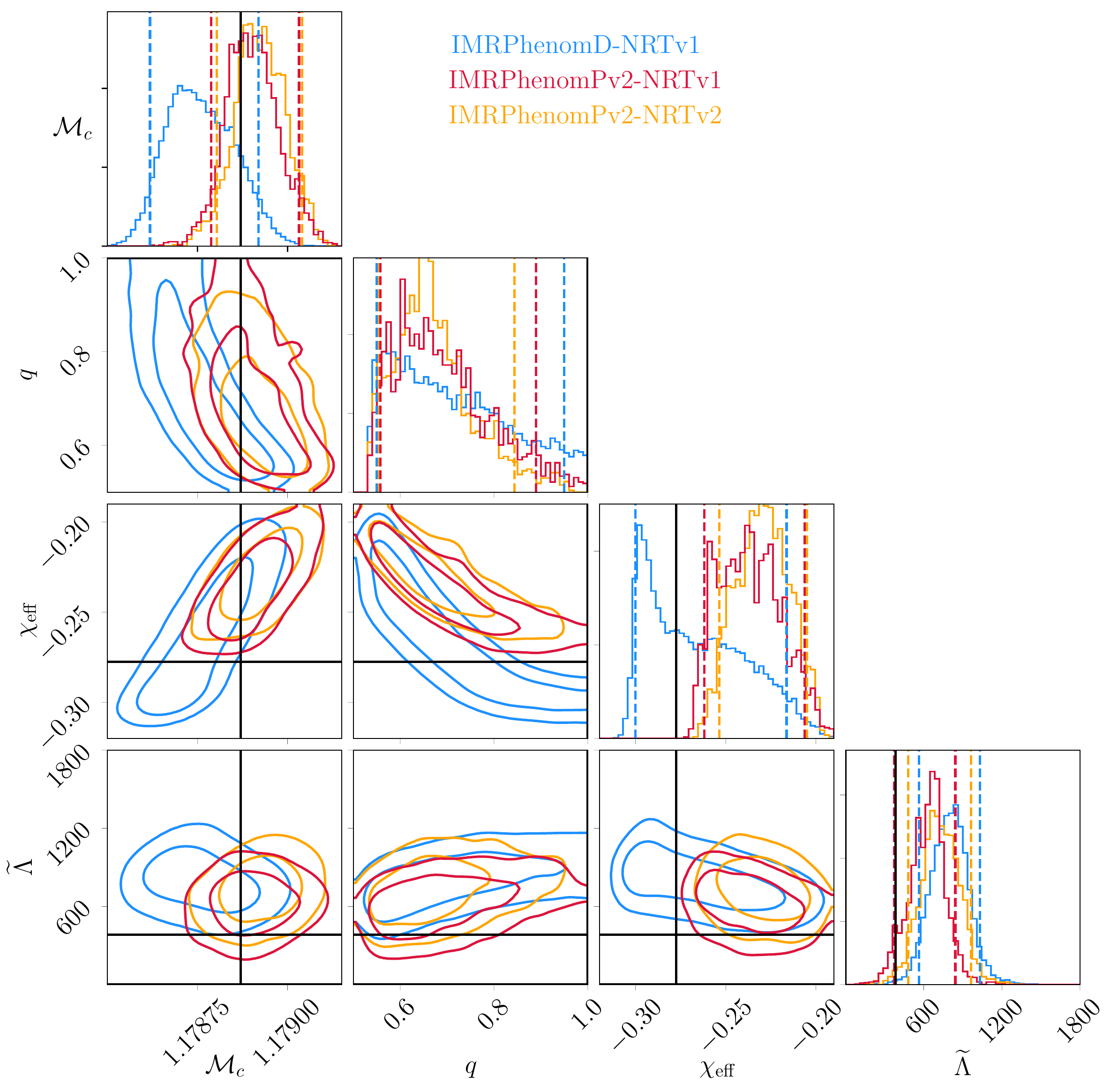}
\caption{Estimated chirp mass, mass ratio, effective spin, and tidal deformability for the 
         \Slydd{} setup. The injected source parameters are marked with a solid black line. 
         Shown contours in the 2D-plots refer to 50\% and 90\% credible intervals, 
         intervals marked in the 1D plots refer to 90\% credible intervals.
         }
\label{fig:PE_sly3dd}
\end{figure*}

%%%%%%%%%%%%%%%%%%%%%%%%%%%%%%%%%%%%%%%%%##
\subsection{Parameter Estimation}
\label{sec:PE}

To understand possible biases during parameter estimation for highly spinning systems, 
we perform an injection study using the LALInference code~\cite{lalsuite,Veitch:2014wba}. We employ the Markov-Chain Monte Carlo (MCMC) algorithm to estimate posterior probability distribution functions; cf.~Figs.~\ref{fig:PE_sly5uu} and  \ref{fig:PE_sly3dd}.
We inject hybrid waveforms starting from 23 Hz. The hybrids are a combination of the updated \SEOBT model to cover the early-inspiral part and the highest resolution NR waveforms\footnote{We do not use the Richardson-extrapolated waveform to avoid any potentially unphysical behaviour of the waveform once convergence is lost after the merger. We employ the methods outlined in Ref.~\cite{Dudi:2018jzn} for hybdridization.}. 
To reduce computational costs, we only perform studies using the \Slyuuh{}   and \Slydd{}  configuration. 
The injected waveforms have an SNR of 32.4 and an inclination of about 0 deg. We perform zero noise injection but assume design sensitivity of Advanced LIGO and Advanced Virgo. 
For the analysis, we assume a uniform prior distribution in the interval $[1 M_\odot, 3 M_\odot]$ for component masses and $[-0.9, 0.9]$ for both dimensionless aligned spins. We only consider non-precessing recoveries. We recover injections with  \IMRPhenomPNRtidal2, \IMRPhenomP1NRtidal, and \IMRPhenomDNRtidal. 

To check the sanity of our runs and to ensure that our pipeline works properly, we inject in addition a \IMRPhenomPNRtidal2 waveform constructed using same parameters as \Slyuuh{} and we recover this injection with the same \IMRPhenomPNRtidal2 model. We label this additional injection as \IMRPhenomPNRtidal2-inj-same and do not expect to see biases for this setup. 

In Fig.~\ref{fig:PE_sly5uu}, we show the results for the \Slyuuh{} setup, where we show the 2D and 1D marginalized probability densities for a subset of the recovered posteriors, namely the chirp mass $\mathcal{M}_c$, the mass-ratio $q$, the effective spin parameter $\chi_{\rm eff}$, and the tidal deformability $\tilde{\Lambda}$. In the 2D contour plots, we show the 50\% and 90\% credible interals. The solid black line marks the injected value.

Through the comparison with \IMRPhenomPNRtidal2-inj-same, 
we find that our employed setup is able to recover the injected parameters reliably and that all injected parameters are recovered within the 90\% confidence interval. 
To the contrary, using our hybrid waveform for the \Slyuuh{} setup, we find much larger discrepancies. Most notably, while obtaining sharp posteriors, the \IMRPhenomDNRtidal model estimates the chirp mass, mass ratio, spin, and tidal deformability wrong. As discussed before, this inaccuracy is introduced by the missing EOS-dependence of the spin-induced quadrupole moment. 
All other waveform approximants recover the injected values (dashed line) within the 90\% credible interval, which, to a large extent, is possible due to generally large uncertainties of the obtained posterior distributions. 

In Fig.~\ref{fig:PE_sly3dd}, we show the results for our injection study for the \Slydd{} setup. 
It is notable that \IMRPhenomDNRtidal performs better than in the previous case, which is consistent with our investigations in Fig.~\ref{fig:compareGW}. However, the model still shows the largest bias in the chirp mass, as well as a wrong recovery of the mass ratio. 
Finally, it is worth pointing out that none of the models is capable of recovering the effective spin and tidal deformability correctly. 
Hence, it is obvious that tidal and spin measurements will be biased noticeably for systems with large anti-aligned spin. 

%%%%%%%%%%%%%%%%%%%%%%%%%%%%%%%%%%%%%%%%%%%%%%%%%%%%%%%%%%%%%%%%%%%%%%%%%%

\section{Conclusion}
\label{sec:conclusion}

Within this work, we have presented high-resolution 
simulations for a set of four different physical configurations
of highly spinning, equal mass binary neutron star systems. 
Every setup has been evolved with a total of four resolutions, 
which allows a precise computation of uncertainties. 
We find that for the highest spinning configurations $\chi=0.58$ and 
$\chi=-0.28$ existing waveform models do not provide an accurate description 
during the late inspiral. 
It is worth pointing out that the dephasing between the \SEOBT model and our numerical-relativity simulations can be reduced by incorporating spin-dependent resonance effects as outlined in Ref.~\cite{Steinhoff:2021dsn}. 
To understand the influence of this disagreement on gravitational-wave analysis, 
we have performed an injection study in which we have tried to recover the source parameters of a hybrid waveform consisting of our high-resolution numerical-relativity data and 
the updated \SEOBT model. Most notably for large anti-aligned spins, estimated tidal 
deformabilities and effective spins are biased and it was not possible to recover the injected 
source parameters. 

Our study shows that, at least for the tested waveform models, 
further development is needed to ensure a reliable interpretation of future gravitational-wave detections 
of possibly highly spinning binary neutron star systems.

%%%%%%%%%%%%%%%%%%%%%%%%%%%%%%%%%%%%%%%%%%%%%%%%%%%%%%%%%%%%%%%%%%%%%%%%%%%%%
\begin{acknowledgments}
%%%%%%%%%%%%%%%%%%%%%%%%%%%%%%%%%%%%%%%%%%%%%%%%%%%%%%%%%%%%%%%%%%%%%%%%%%%%%%%
  It is a pleasure to thank Sebastiano Bernuzzi 
  and Rossella Gamba for helping in constructing TEOBResumS waveforms, 
  Ajit Kumar Mehta for helping in setting up of parameter estimation runs, 
  and Antoni Ramos-Buades for helpful comments with respect to the manuscript. 
  R.D.\ thanks Masaru Shibata for helpful discussions throughout the duration of this project. 
  T.D.\ acknowledges financial support through the Max Planck Society. 
  R.D.\ and B.B.\ were supported in part by DFG grant BR 2176/5-1.
  W.T.\ was supported by the National Science Foundation under grants PHY-1707227 and PHY-2011729.
  Computations were performed on HAWK at the High-Performance Computing 
  Center Stuttgart (HLRS) [project GWanalysis 44189], 
  on SuperMUC\_NG of the Leibniz Supercomputing Centre 
  (LRZ) [project pn29ba], and on the Minerva cluster of the 
  Max Planck Institue for Gravitational Physics.

  %
%%%%%%%%%%%%%%%%%%%%%%%%%%%%%%%%%%%%%%%%%%%%%%%%%%%%%%%%%%%%%%%%%%%%%%%%%%%%%
\end{acknowledgments}

\bibliography{paper20210823.bbl}

%%%%%%%%%%%%%%%%%%%%%%%%%%%%%%%%%%%%%%%%%%%%%%%%%%%%%%%%%%%%%%%%%%%%%%%%%%%%%%%
\end{document}